# Theoretical Investigation of the Periodic Anderson Hamiltonian of Samarium Hexaboride

*Partha Goswami and Udai Prakash Tyagi*

*D.B.College, University of Delhi, Kalkaji, New Delhi-110019, India*

*Email of the corresponding author: physicsgoswami@gmail.com*

Email of the second author: uptyagi@yahoo.co.in

**Abstract**

The periodic Anderson Hamiltonian of the bulk samarium hexaboride is investigated in this article assuming the presence of ferromagnetic impurities (FM). The problem of large on-site electron-electron repulsion is reformulated in terms of a holonomic constraint using slave-boson technique. The model analysis yields the effective mass of electron and the possibility of the quantum anomalous Hall phase. Upon using the Fu-Kane-Mele formalism, it is indicated that the surface Hamiltonian without FM may correspond to a strong topological insulator.

## 1.Introduction

The strong correlation effects, the hybridization involving Sm 5d and $4f$ electrons, and diverse surface conditions render investigation of the problem of Samarium hexaboride ($SmB_6$) extremely complicated **[1-18]** to deal with. The compound, with a high-temperature correlated metallic phase, transforms into Kondo insulator below 60 K **[8].** The investigation of the problem presented here is based on the periodic Anderson model (PAM) extended using the slave-boson (SB) formalism **[3, 17,18]**. The hybridization term (HT) of the model is responsible for topological dispensation of the compound. Only the lowest-order cubic harmonics have been taken into account to represent HT. The communication involves scrutiny of two issues concerning this 3D topological Kondo Insulator (TKI) **[1-3]**. These are explained below in brief.

It will be shown that the proximity to the ferromagnetic magnetic (FM) impurities, which breaks time reversal symmetry (TRS), leads to the possibility of the quantum anomalous Hall (QAH) effect (observed usually in 2D systems) in the insulating bulk for certain parameter window(s) as the Chern number will be shown to have integer values. The exercise is motivated by the work of Kim et al. **[4]** where it was reported that the compound $SmB_6$ is sensitive to the presence of impurities. The possible onset of this novel phase happens in the bulk notwithstanding the absence of band-crossing at discrete nodes **[19]**. The reason is that along paths $\mathcal{P}$ connecting high-symmetry points in the three-dimensional Brillouin zone (BZ) of $SmB_6$ – a cubic system – the wave vector component ($ak_z$) is either zero, or $\pi$. Consequently, an investigation focused along $\mathcal{P}$ is effectively $ak_z$-independent. Thus, without loss of generality, one can then think of the (PAM) Hamiltonian involving the wave vector components $(ak_{x,y}, 0/\pi)$ as a two-dimensional Hamiltonian, which may either be topologically trivial or non-trivial in the absence of FM.

It has been suggested [1,2], as well as there is mounting evidence [20-22] during the past several years, that the TKI SmB$_6$ possesses non-trivial topology. This has generated great deal of excite-ment in the condensed matter physics community and it still remains a matter of intense debate [23, 24]. The supporting evidence for the non-triviality scenario [20-22] notwithstanding, there is no jury-verdict regarding the nature of the (bulk and) surface states of SmB$_6$ [23–25]. Upon using the Anderson model with SB formalism (FM exchange energy $M = 0$) around $\bar{\Gamma}$ (0,0) and $\bar{X}$ ($\pi$, 0), obtained by the projection of the $X$ points of the bulk BZ in the (001) surface BZ, it is indicated here in the Fu-Kane framework [26-28] that the topological invariant $Z_2 = -1$. Thus, the novelty of the present work lies in the fact that it is able to show the compound SmB$_6$ possesses a strongly topological surface state (TSS) within the frame work of PAM extended by the SB formalism in the absence of FM. The possibility of QAH phase in the presence of FM, as already mentioned above, is another novel aspect of the work. It may be mentioned that signature of two-dimensional Fermi surfaces on (100) and (101) surface planes supporting TSS were obtained in the quantum oscillation experiments of Li et al. [29].

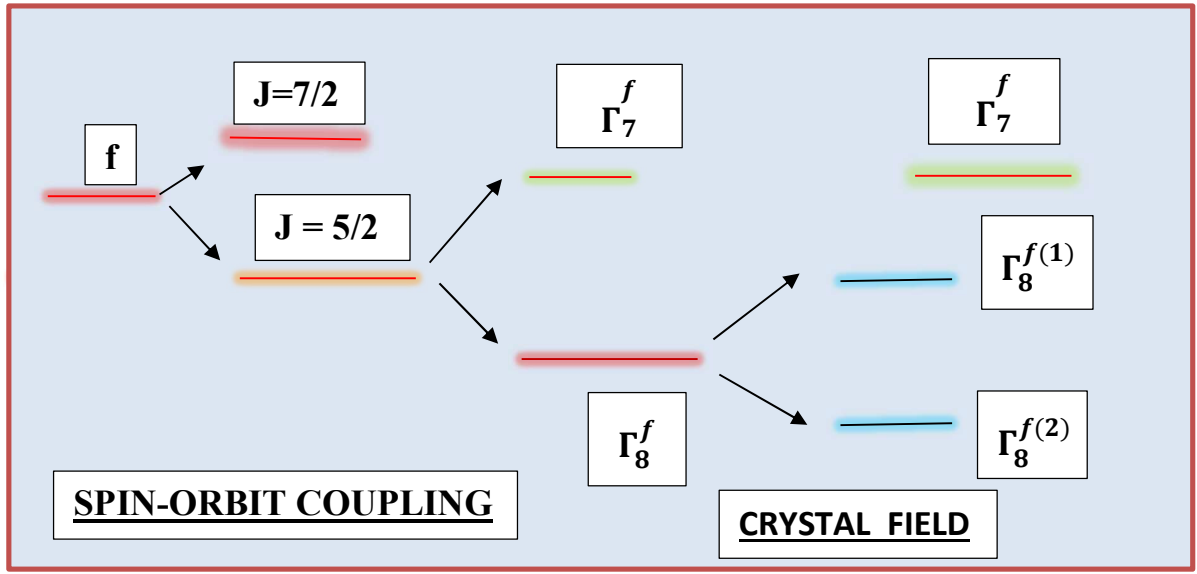

**Figure 1.** A representation of the of the evolution of energy levels of the *f*-states in SmB$_6$ owing to mediation of the spin-orbit coupling and the crystal field.

It was found earlier within the local density approximation + Gutzwiller method incorporating a Green's function scheme [16,30,31] that the states near the Fermi energy in SmB$_6$ are formed by the Sm 5d electrons (exhibiting $t_{2g}$ and $e_g$ symmetry) and the Sm 4$f$ electrons. The *f*-states are split into $J = 7/2$ and $J = 5/2$ states by spin-orbit coupling. The $J = 5/2$ state is slightly below the Fermi energy $E_F$. It splits into $\Gamma_7^f$ doublet and a $\Gamma_8^f$ quartet due to presence of the crystal field (see

Figure 1). The $\Gamma_8^f$ quartet further splits into $\Gamma_8^{f(1)}$ and $\Gamma_8^{f(2)}$ doublets. The fermionic operator creating a localized doublet state at a site **r**, associated with the crystal field effect and spin-orbit coupling, will be denoted by $f_{m_j}^\dagger(\mathbf{r})$. In the case of SmB$_6$, the lowest-lying doublet is $|\Gamma_8^{f(2)}\rangle = |m_j = \pm\frac{1}{2}\rangle$, where $m_j = \pm\frac{1}{2}$ (or, $m_j = \uparrow, \downarrow$) is a pseudospin quantum number, corresponding to the two possible values of the projection of total angular momentum in the lowest-lying state. The hybridization between the $d$-electron operator $d_\tau^\dagger(\mathbf{r})$ (where $\tau (=\uparrow, \downarrow)$ is the spin index) and $\Gamma^f$ operator $f_\alpha(\mathbf{r}')$ (where $\alpha = m_j$ is the internal quantum number of $f$ orbitals prior to the hybridization) are supposedly responsible for the opening of a larger gap $\Delta(\sim 20\ meV)$ and a smaller one $\Delta'(\sim 3 meV)$ [30 − 34]. These complexities have been included in the earlier reports **[30-34]** as well as in the present one to a certain degree. The corresponding energy dispersion will be obtained in section 2.

The periodic Anderson model **[3,17,18]**, which is the theoretical description of a Kondo insulator **[35, 36]** at the minimalistic level, is discussed below. The model describes hybridization ($V$) between the conduction $d$-electrons (even parity) and localized, strongly-interacting $f$-electrons (odd parity) as already mentioned. The conduction electrons take care of the Kondo screening **[37]** of a localized magnetic moment. The nearest-neighbor (NN) hopping energy of these electrons $t_{d_1}$ $\sim 25\ meV$. The NN hopping ($t_{f_1}$) of $f$-electrons may be assumed to be smaller than $t_{d_1}$. The correlations ($U_f$) of $f$-electrons is stronger on the surface than on the bulk. The topological order is due to $V$ being comparable to the hopping energy of $d$ – electrons, that is $V \sim t_{d1} \sim 15 − 20\ meV$. In the opposite case, one obtains the Kondo lattice Hamiltonian**[38]**. The latter describes interaction between spins of localized and conduction electrons leading to Kondo singlet formation. This minimalistic SmB$_6$ (bulk) model **[3,17,18,39]**, with $U_f \gg t_{d1}$ and $V \sim t_{d1}$, captures essential physics of TKI, such as the bulk and surface band-structure topology as shown in this report. The constraint $U_f \gg t_{d_1}$ imposes a non-holonomic constraint, viz. the exclusion of the double occupancy, which is very difficult to manipulate. The slave boson (SB)- framework provides a platform to reformulate this nonholonomic constraint into a holonomic constraint.

The paper is organized as follows: In section 2, the effective mass of electrons and the bulk band spectrum in the SB formalism are obtained. Next, the Chern number is calculated to show the possibility of the quantum anomalous Hall (QAH) phase in the presence of FM. The surface state Hamiltonian is obtained in section 3 around $\bar{\Gamma}$ (0,0) point in the absence of FM. The Z$_2$ invariant is also calculated within the Fu-Kane-Mele framework **[26-28]** in this section. The highlights of the present work and the associated complexities are discussed in section 4. The paper ends with very brief concluding remarks in section 5.

2. **Model and Method**

The minimalistic tight-binding Hamiltonian (PAM) **[2,3,17]** for the compound SmB$_6$ is given by

$$H_{PAM} = \sum_{k,\tau} E_k^d \, d^\dagger_{k,\tau} d_{k,\tau} + \sum_{k,\alpha} E_k^f \, f^\dagger_{k,\alpha} f_{k,\alpha} + \sum_{r,r'} \sum_{\alpha,\tau} \{F_{\alpha,\tau=\uparrow,\downarrow}(r,r') \, d^\dagger_\tau(r) f_\alpha(r') + \text{H.C.}\}$$

$$+ U_f \sum_r \sum_{\alpha,\beta} f^\dagger_\alpha(r) f_\alpha(r) f^\dagger_\beta(r) f_\beta(r) \qquad (1)$$

on a simple cubic lattice with the lattice constant $a$. The dispersions $E_k^d = -[2t_{d1} c_1(k) + 4t_{d2} c_2(k) + 8t_{d3} c_3(k)]$ and $E_k^f = [-\epsilon_f - 2t_{f1} c_1(k) - 4t_{f2} c_2(k) - 8t_{f3} c_3(k)]$ for $d$– and $f$– electrons, respectively, capture spread of atomic orbitals into bands. The terms $(t_{d1}, t_{f1})$, $(t_{d2}, t_{f2})$, and $(t_{d3}, t_{f3})$, respectively, are the NN, NNN, and NNNN hopping parameters for $d$ and $f$ electrons ($\epsilon_f$ is the onsite energy of the $f$ electrons), $c_1(k) = \sum_{j=(x,y,z)} \cos(ak_j)$, $c_2(k) = \sum_{i \neq j=(x,y,z)} \cos(ak_i) \cos(ak_j)$, and $c_3(k) = \prod_j \cos(ak_j)$. The annihilation operators for $d$– electron in momentum($k$)-space is represented by $d_{k,\tau=\uparrow,\downarrow}$. The symbol $U_f$ stands for the on-site repulsion among the $f$- electrons. Furthermore, prior to the hybridization represented by the term $F_{\alpha,\tau=\uparrow,\downarrow}(r,r')$, one may regard the spin $\tau$ as a good quantum number as the model given by (1) neglects the spin-orbit coupling of $d$ electrons. Furthermore, the operator $f_\alpha(r)$ is related to the corresponding momentum space operator $f_{k,\alpha}$ by a simple Fourier transform $f_{k,\alpha} = \sum_r \exp(-ik \cdot r) f_\alpha(r)$. Similarly, the Fourier transform of the hybridization parameter is $F_{\alpha,\tau}(k) = \sum_{r,r'} \exp(-ik \cdot (r-r')) F_{\alpha,\tau=\uparrow,\downarrow}(r,r')$. Since the $f$- and $d$-states have different parities, the momentum-dependent form-factor $F(k)$ involved in the third term in (1) must be odd. This is required in order to preserve time reversal symmetry (TRS), as the term involves coupling of the pseudo-spins with the physical spin of the electron. Therefore, one can write $F_{\alpha,\tau}(k)$ as $2V(s(k) \cdot \tau')$, where $V$ is the constant hybridization amplitude, $s(k) = (\sin(ak_x), \sin(ak_y), \sin(ak_z))$, and $\tau' = (\tau_x, \tau_y, \tau_z)$ are the Pauli matrices in spin space.

A brief account of the SB-framework based extension of the mean-field-theoretic version of PAM [3,17,18] for a (topological) Kondo insulator is given below: As the first step of preliminaries, one makes the replacement $f_\alpha(r) \to c_\alpha(r) b^\dagger(r)$ where $f_\alpha(r)$ is annihilation operator for an $f$-electron, $c_\alpha(r)$ is a pseudo-spinful fermion annihilation operator, and $b^\dagger(r)$ is a spinless slave-boson creation operator at a site $r$. The introduction of slave boson gives rise to extension in the Hilbert space of the system. The complications associated with the large on-site repulsion ($U_f \gg t_{d1}$) between the $f$-electrons, implying no double occupancy of a site, is then conveniently circumvented by imposing the holonomic constraint $[\sum_\alpha \langle c^\dagger_\alpha(r) c_\alpha(r) \rangle + \langle b^\dagger(r) b(r) \rangle] = 1$. As the next step, in order to simplify the problem, one assumes no spatial dependence of the boson operators and, in addition, replace them by their expectation value $(\langle b^\dagger(r) b(r) \rangle \to |b|^2$. Here $b$ may be complex as the density distribution of the Bose-condensate is represented by a wavefunction with a well-defined amplitude and phase. Since one needs to recover the physical subspace of the extended Hilbert space, thus additionally $\sum_\alpha \langle f^\dagger_\alpha(r) f_\alpha(r) \rangle = |b|^2 \sum_\alpha \langle c^\dagger_\alpha(r) c_\alpha(r) \rangle$. The implication is whereas the pseudo-spinful fermions undergo hybridization with the itinerant $d$-electron, the spinless bosons form the SB condensate. The constraint equation now appears as $|b|^{-2}\sum_\alpha \langle f_\alpha^\dagger(r)f_\alpha(r)\rangle = 1 - |b|^2$. With fluctuations of $(b^\dagger(r)b(r))$ frozen, this holonomic constraint is then imposed in a mean-field fashion in ref.[3,17] using a Lagrange multiplier λ. For the formation of Kondo singlet states, between $f$ and $d$ fermions, one requires $N_c = N_d$, where $N_c$ ($N_d$) correspond to the number of $f$- ($d$-) fermions. The auxiliary chemical potential $\xi$ enforces the fact that there are equal number of $f$ and $d$ fermions. The chemical potential $\mu$ of the fermion number, which is a free parameter, enforces that $[\sum_{k,\alpha}\langle c_{\alpha,k}^\dagger c_{\alpha,k}\rangle + \sum_{k,\tau}\langle d_{k,\tau}^\dagger d_{k,\tau}\rangle]$ is the total number of fermions. The non-free parameters are (λ, $|b|$ and $\xi$) [3,17,18,39]. It is found that $\lambda = -6t_{f1} + 6b^2 t_{f1}$, and $\xi = -3t_{d1} + 3t_{f1}$, where $t_{f1} < 0$ ($t_{f1} > 0$) for the insulating (conducting) bulk. The numerical value of $|b|^2$ depends on the choice of the values of the hopping and the hybridization parameters in Eq.(1). The estimated value of this parameter varies from 0.50 to 0.90. As will be shown below, the estimate is in reasonable agreement with experiment [40].

One of the actualizations of the strongly correlated electron system is that the effective mass $m^*$ of electron is expected to be larger than $m_e$, where $m_e$ is the bare electron mass. Since the bulk SmB$_6$ is a mixed valent insulator with strong correlations between $f$-electrons, it is interesting to estimate the ratio $R = m^*/m_e$. This is important as well in view of the scanning-tunnelling spectroscopic experimental result [40] where $R$ was found to be 100-1000. For this purpose, the constraint-equation is considered once again. It appears as $\sum_{\alpha, ak<ak_F}\langle f_{k,\alpha}^\dagger f_{k,\alpha}\rangle = N(|b|^2 - |b|^4)$ where $N$ is the total number of unit cells. In the low-temperature limit, the expectation value above needs to be replaced by the usual Heaviside step function. The multiplication of the Fermi momentum $k_F$ with the lattice constant $a$ yields a dimensionless quantity. The integration on the left-hand side is non-trivial, as assigning a single value to the Fermi wave vector ($ak_F$) of an anisotropic (lattice) band structure, whose Fermi surface (FS) may not be a sphere, is inappropriate. On the experimental front, it is known [41] that the mapping of the quantum oscillation (QO) frequencies of a crystalline sample with respect to the tilt angle of the applied magnetic field enables one to acquire information about the charge carriers and the shape and sizes of FS. In the case of pristine pure single crystals of SmB$_6$ at low temperature and high magnetic field, the experimental measurements concerning torque magnetometry [41] has unveiled QO oscillation frequencies being similar to a large three-dimensional conduction electron FS [41, 42]. This unusual observation is possibly due to slow fluctuations between a collectively hybridized insulating state and an unhybridized state in which the conduction electrons form a solely conduction electron FS. Upon taking cue from this experimental result, FS of the system is assumed to be a sphere in the first approximation. This leads to the equation

$$|b|^2 = \left(\frac{1}{2}\right)\left[1 \pm \sqrt{\left(1 - \frac{\left(\frac{32\pi}{3}\right)\hbar^3 a^3 k_F^3}{\hbar^3}\right)}\right], \qquad (2)$$

The equation is basically the one for finding $ak_F$. Upon using the relation $\hbar a k_F = m^* a v_F$, $|b|^2 \approx 0.50 - 0.90$, and the value of the Fermi velocity $(v_F) \sim 0.5 \times 10^3 \frac{m}{s}$ [27], we obtain $R \sim 150$. This value of the ratio $R$ is in reasonable agreement with refs. [40, 43].

The presence of the ferromagnetic magnetic (FM) impurities in the system Hamiltonian is an additional feature. The impurities are assumed to be interacting with $d$ electrons only. The magnitude of the impurity spin $|S| > 1$ could be absorbed into the ferromagnetic coupling constant $J$ and a term involving $M = |J||S|$ could be introduced. The constant term $\lambda N_c (|b|^2 - 1)$ in the Hamiltonian in [3,39] is omitted below. In the basis $(d_{k,\uparrow}^\dagger\ d_{k,\downarrow}^\dagger\ |b|c_{k,\uparrow}^\dagger\ |b|c_{k,\downarrow}^\dagger)$, the mean-field theoretic SB Hamiltonian matrix could now be written as

$$h_{extended}(k, |b|, M) = \frac{\widetilde{E_k^d}(\mu)}{2}(\mathbb{I} + \gamma^0) + \frac{\widetilde{E_k^f}(\mu|b|)}{2}(\mathbb{I} - \gamma^0) + 2C_\mu \Sigma^\mu + \vartheta_\mu \gamma^0 \gamma^\mu. \qquad (3)$$

Here $\mathbb{I}$ is the identity matrix, $C_\mu = (0, 0, M)$, $\Sigma^\mu = \left(\frac{1}{2}\right) \epsilon^{\mu\nu\rho} \sigma_{\nu\rho}$, $\sigma_{\nu\rho} = \left(\frac{i}{2}\right)[\gamma_\nu, \gamma_\rho]$, and $\vartheta_\mu = (\vartheta_x, \vartheta_y, \vartheta_z)$. These components, respectively, are $\vartheta_x = 2V|b|\sin a k_x$, $\vartheta_y = 2V|b|\sin ak_y$, and $\vartheta_z = 2V|b|(\sin ak_z)$. The hybridization amplitude $V$ is renormalized by $|b|$. The additional term $2C_\mu \Sigma^\mu$ breaks TRS. The Dirac matrices $(\gamma^0, \gamma^1, \gamma^2, \gamma^3, \gamma^5)$ in contravariant notations are $\gamma^0 = \tau^z \otimes I^{2\times 2}$, $\gamma^j = i\tau^y \otimes \tau^j$, $\tau^i$ are Pauli matrices, $j = 1,2,3$, and $\gamma^5 = i\gamma^0 \gamma^1 \gamma^2 \gamma^3$. The renormalized dispersion of $d$- and $f$-electrons, respectively, are given by $\widetilde{E_k^d}(\mu) = -\mu - \xi + E_k^d$ and $\widetilde{E_k^f}(\mu, |b|) = -\mu + \xi + |b|^2 E_k^f + \lambda$. The hybridization bandgap $\Delta (\sim 20\ meV)$ is controlled by the energy scale $|b|V$ which must be of the same order of magnitude as $\Delta$. As one can see from the SB protocol version of PAM given by Eq. (3), when $|b| \to 0$, the system is a non-interacting lattice gas mixture of itinerant $d$- and non-hopping $f$- fermions with no topological dispensation. The system shows the bulk metallic ($t_{f_1} > 0$) as well as the bulk insulating ($t_{f_1} < 0$) phases. The negative sign of $t_{f_1}$ is also necessary for the band inversion, which induces the topological state [3]. Throughout the paper, we choose $t_{d_1}$ to be the unit of energy. The eigen-values $E_n$ of the Hamiltonian matrix $h_{extended}(k, |b|, M)$ in (3) are given by

$$E_n(s, \tau, k, |b|, M, V) = \frac{(\widetilde{E_k^d}(\mu) + \widetilde{E_k^f}(\mu, |b|) + \tau M)}{2} + s \left[\frac{(\widetilde{E_k^d}(\mu) - \widetilde{E_k^f}(\mu, |b|) + \tau M)^2}{4} + \vartheta_x^2 + \vartheta_y^2 + \vartheta_z^2\right]^{1/2} \qquad (4)$$

where $n = 1,2,3,4$, $\tau = \pm 1$ is the spin/pseudo-spin index and $s = \pm 1$ is the band-index. The bulk eigenstates corresponding to the eigenvalues in (4) are given in Appendix A.

The high symmetry points (HSPs) of the bulk Brillouin zone (BZ) $\Gamma(0,0,0)$, $X\{(\pi, 0,0), (0, \pi, 0), (0,0,\pi)\}$, $M\{(\pi, \pi, 0), (\pi, 0, \pi), (0, \pi, \pi)\}$ and $R(\pi, \pi, \pi)$ are to be considered for plotting the single-particle spectra given by (4). The $X$ and $M$ points, three for each of them, are equivalent for symmetry reasons. The projection of the $X$ points in the (001) surface BZ gives $\overline{\Gamma}$ (0,0) and $\overline{X}$ ($\pi$, 0) will be referred to in section 3 for the construction of two-dimensional Hamiltonian. We have plotted the bulk band energies of the system in Eq.(4) for $M \neq 0$ in Figure 2. The numerical values of the parameters used in the plots are $t_{d_1} = 1.0000$, $|t_{f_1}| = 0.4835$, $t_{d_2} = 0.1$, $t_{f_2} = 0.1$, $t_{d_3} = 0.01$, $t_{f_3} = 0.01$, $\epsilon_f = -0.02$, V= 0.16, $|b|^2 = 0.83$, $\mu = 0$, and $M = 0.50$. A 3D diagrammatic representation of the four bands (given by Eq.(4)) with spectral gap, with $t_{f_1} < 0$, and $M \neq 0$, is shown in Figure 2(a). In 2(b) and 2(c), we have the plots of the same four bands as a function of the wave number component $ak_x$ along the paths $\mathcal{P}: \bar{M}(-\pi, 0, \pi) - X(0,0,\pi) - M(\pi, 0, \pi)$ and $\bar{R}(-\pi, \pi, \pi) - M(0, \pi, \pi) - R(\pi, \pi, \pi)$, respectively, with $t_{f_1} < 0$ (insulating bulk), and $M \neq 0$. The Fermi energy is represented by the horizontal line. The importance of the spectral gap in these figures cannot be overstated for the integer value of the Chern number $\mathcal{C}$ to exist. That is, if the gap does not exist, calculating the Chern number becomes infructuous. In Figure 2(d), the plot of the four bands along the path $\bar{M}(-\pi, 0, \pi) - X(0,0,\pi) - M(\pi, 0, \pi)$ is for the metallic bulk case($t_{f_1} > 0$) where $\mathcal{C}$ may have non-integer values. A 3D representation of the four bands showing no spectral gap (as $t_{f_1} > 0$) is shown in Figure 2(e). The spectra of the 3D system here display no band-crossing feature at discrete nodes as reported in 3D (Weyl) systems **[19]**; this was found to be essential for the integer chern number $\mathcal{C}$ to exist. However, as has been explained in section 1, along the path(s) $\mathcal{P}$ connecting high-symmetry points in the three-dimensional Brillouin zone (BZ) of SmB$_6$ we have the $ak_z$-independence. This means along these paths, effectively, we have a 2D system.

The anomalous Hall conductivity (AHC) is given by $\sigma_{AH} = -(\frac{e^2}{\hbar}) \sum_j \int_{BZ} \frac{d^2k}{(2\pi)^3} g(E_j(k) - \mu) \Omega_j^z(k)$, where $\mu$ is the chemical potential of the fermion number, $j$ is the occupied band index, $g(E_j(k) - \mu)$ is the Fermi-Dirac distribution function and $\Omega_j^z(k)$ is the z-component of the Berry curvature (BC) for the $j$ th band. To obtain AHC, the Berry curvature is calculated using the Kubo formula

$$\Omega_j^z(k) = -2\hbar^2 [Im \sum_{i \neq j} (E_j(k) - E_i(k))^{-2} \langle j, k|\widehat{v_x}|i, k\rangle \langle i, k|\widehat{v_y}|j, k\rangle]. \tag{5}$$

Here $\boldsymbol{k}$ is the Bloch wave vector, $E_j(k)$ is the band energy, $|j, k\rangle$ are the Bloch functions of a single band. The operator $\widehat{v_x}$ represents the velocity in the $x$ direction. For a system in a periodic potential and its Bloch states as the eigenstates, in view of the Heisenberg equation of motion $i\hbar \frac{d\hat{x}}{dt} = [\hat{x}, \hat{H}]$, the identity $\langle m, \boldsymbol{k'}|\widehat{v_\alpha}|n, \boldsymbol{k}\rangle = \left(\frac{1}{\hbar}\right)\left(E_j(\boldsymbol{k'}) - E_i(\boldsymbol{k})\right)\left\langle i, \boldsymbol{k'}|\frac{\partial}{\partial k_\alpha}|j, \boldsymbol{k}\right\rangle$ is satisfied. Upon using this identity, one obtains AHC in the zero temperature limit as $\sigma_{AH} = C\left(\frac{e^2}{\hbar}\right)$ where $C = \sum_j C_j$, $C_j = \int \int_{BZ} \Omega_j^z(k) \frac{d^2k}{(2\pi)^2}$. The z-component of the Berry-curvature(BC) is

$$\Omega^z(k) = \sum_j \left(\frac{\partial A_{j,y}}{\partial k_x} - \frac{\partial A_{j,x}}{\partial k_y}\right) = -2 \sum_j Im \left\langle \frac{\partial u_{j,k}}{\partial k_x} \Big| \frac{\partial u_{j,k}}{\partial k_y} \right\rangle \tag{6}$$

where $|u_{j,k}(k)\rangle = |j,k\rangle$. The Berry curvature is the analogue of the magnetic field in momentum-space while the Berry connection $A_\alpha(k)$ acts as a vector potential; that is, $\nabla_k \times A_\alpha(k) = \Omega_\alpha(k)$. The result in Eq. (6) has been used to calculate BC. The high symmetry points (HSPs) of the bulk Brillouin zone (BZ) are given above. In line with the bulk-boundary correspondence, upon considering these points and the eigenvectors of the bulk system in the Appendix A, BC is calculated analytically ensuring that all occupied levels are taken in. Next, on integrating BC on a k-mesh-grid (with appropriate resolution) of the Brillouin zone, the intrinsic AHC has been obtained. The 'Mat-lab' package has been used for this purpose. The contour plots of BC are given in Figure 3(a) and 3(b). The chern number calculated in the two cases are **(a)** $C = -2.0043 \sim -2$, and **(b)** $C = 2.0073 \sim +2$. One, thus, finds the possibility of the QAH phase for the bulk system. The numerical values of the parameters used in the plots are $t_{d_1} = 1$, $t_{f_1} = -0.5$, $t_{d_2} = 0.1$, $t_{f_2} = 0.1$, $t_{d_3} = 0.01$, $t_{f_3} = 0.01$, $\epsilon_f = -0.02$, $\mu = 0$, and $|b|^2 = 0.83$. The additional parameter values in **(a)** $ak_z = \pi$, $t_{f_1} = -0.515$, $V = 0.50$, and $M = 0.20$, while in **(b)** $ak_z = 0$, $t_{f_1} = -0.491$, $V = 0.620$, and $M = 0.10$. A positive (negative) Chern number implies that the unidirectional edge waves propagate clockwise (anti-clockwise) with respect to the z-axis.

### 3. $Z_2$ invariant for M = 0

The well-known Fu and Kane methodology (FKM) **[26-28]** is to be used below for calculating the $Z_2$ invariant when $M = 0$. The method is based upon the eigenvalues of the parity operator. The FKM based treatment will be presented towards the end of this section after obtaining the surface band spectrum near $\bar{\Gamma}$ (0,0) point of the (surface) BZ using evanescent wave method (EWM). For surface spectrum near the $\bar{X}$ $(\pi, 0)$ point, one may use the similar procedure. An EWM, similar in spirit to that in ref.**[44],** will be followed now in order to obtain the surface state Hamiltonian $(h_{\text{surface}}(k_x, k_y, \mu, |b|, M = 0)$. It is assumed that the plane surface $z = 0$ relates to the length and the breadth of the compound sample. As the first step of EWM, since $ak_z$ is not a good quan-tum number, one makes the replacement $ak_z \rightarrow -ia\,\partial_z$ in the terms involving $\sin(ak_z)$ and $\cos(ak_z)$ in (3) given in section 2. By using the Taylor expansion, one makes these terms appear as polynomial operators $f(\partial_z)$. As the second step, one assumes that the states of the Hamiltonian are quasi-localized within the surface z = 0, and of the form $Ae^{-qz}|u_n(k_x, k_y)\rangle$ for z > 0 and $Ae^{qz}|u_n(k_x, k_y)\rangle$ for z < 0, where $|u_n(k_x, k_y)\rangle$ is the eigenstate of $n^{th}$ eigenenergy of the surface Hamiltonian (see Appendix A). The evanescent states are, thus, simply decaying for z > 0 and z < 0 if A is constant. For A = $\cos(qz)$, as in **[44]**, the states will be oscillatory in space with decaying amplitude. The last step is to use the substitution rule $f(\partial_z)e^{qz} = f(q)e^{qz}$. For A = $\cos(qz)$, the exponential-shift rule $f(\partial_z)e^{qz}u = e^{qz}f(\partial_z + q)u$ may be used. The constant A will be assumed to be one for simplicity. Here '$q$' is a wavenumber such that $q^{-1} \sim d$ where $d$ is a depth introduced to facilitate the estimation of the surface state penetration. If the restriction that the surface state possesses the value $d \sim$ 5-10 nm is put, one obtains $aq \sim$ 0.08 – 0.04. This ensures that the decaying term $e^{-q|z|} \sim e^{-1}$ for $|z| \sim$ 5-10 nm. This value of the penetration depth is comparable to that of $Bi_2Se_3$. On the other hand, for the lower value $d \sim$ 0.13(0.4-0.5) nm, one obtains $aq \sim \pi\ (\sim 1)$. Consequently, $e^{-q|z|}$ will be nearly $e^{-1}$ for very lower value of $|z|$. On a quick side note, it needs mentioning that, for $d \sim$ 0.13 nm, the Hamiltonian near $\bar{\Gamma}$ (0,0) point could be written as

$$h_{\text{surface}}(k, \mu, |b|, M = 0) = \begin{pmatrix} \mathfrak{b}_+ = \mathfrak{b}((k_x, k_y, q, \mu, |b|) & 0 \\ 0 & \mathfrak{b}_- = \mathfrak{b}^*(-k_x, -k_y, q, \mu, |b|) \end{pmatrix} \quad (7)$$

in the basis $(d_{k,\uparrow}\ |b|c_{k,\downarrow}\ \ d_{k,\downarrow}\ \ |b|c_{k,\uparrow})^T$. Here $\mathfrak{h}_+ = (\epsilon_0(k,\mu,|b|))\tau_0 + \boldsymbol{n}(k_x,k_y,|b|)\cdot\boldsymbol{\tau}$, the two blocks $(\mathfrak{h}_+,\mathfrak{h}_-)$, characterized by the pseudo-spin indices $(+,-)$, are related to each other by time reversal symmetry (TRS), and $\boldsymbol{n}(k_x,k_y,q,|b|) = (\vartheta_x,\ \vartheta_y,\epsilon(k_x,k_y,\mu,|b|))$. Equation (7) corresponds to Qi-Wu-Zhang (QWZ) model**[45-47]**. As shown by these authors, the model corresponds to the quantum spin Hall (QSH) state. This state, however, may not be realizable here in view of $a \sim 0.4\ nm$. So, we turn our attention to case where the penetration depth $d \sim$ 5-10 nm. In this case the two pseudo-spin blocks get coupled. With these preliminaries, the Hamiltonian $h_{\text{surface}}(k_x,k_y,\mu,|b|,q)$ near $\bar{\Gamma}$ (0,0) point may be written as

$$h_{\text{surface}}(k,\mu,|b|,q) = (\epsilon_0(k,\mu,|b|))\sigma_0\otimes\tau_0 + \vartheta_x\ \sigma_z\otimes\tau_x + \vartheta_y\ \sigma_z\otimes\tau_y + \epsilon(k,\mu,|b|)\ \sigma_0\otimes\tau_z$$

$$+ \left(\frac{1}{2}\right)[-i\ \vartheta_{z0}\ (\sigma_x + i\sigma_y)\otimes\tau_x + i\ \vartheta_{z0}\ (\sigma_x - i\sigma_y)\otimes\tau_x], \quad (8)$$

$$\epsilon_0(k,\mu,|b|,q) = \frac{(\widetilde{E_k^d}(\mu,q)+\widetilde{E_k^f}(\mu,|b|,q))}{2}, \epsilon(k,\mu,b) = \frac{(\widetilde{E_k^d}(\mu,q)-\widetilde{E_k^f}(\mu,|b|,q))}{2}, \vartheta_{z0} = 2V|b|\sin aq. \quad (9)$$

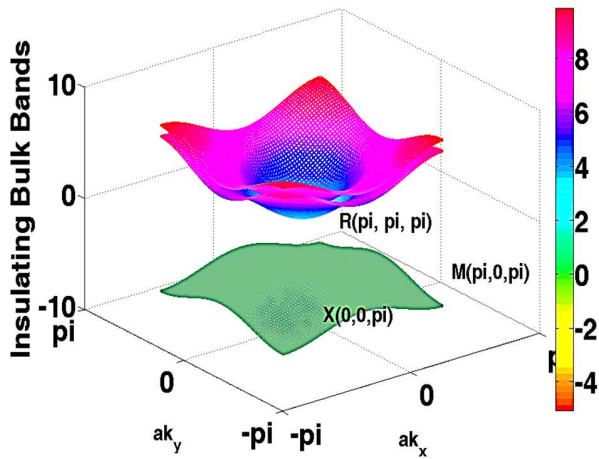

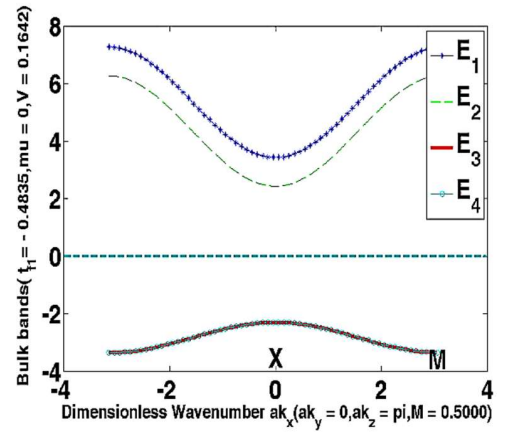

(a)                                                (b)

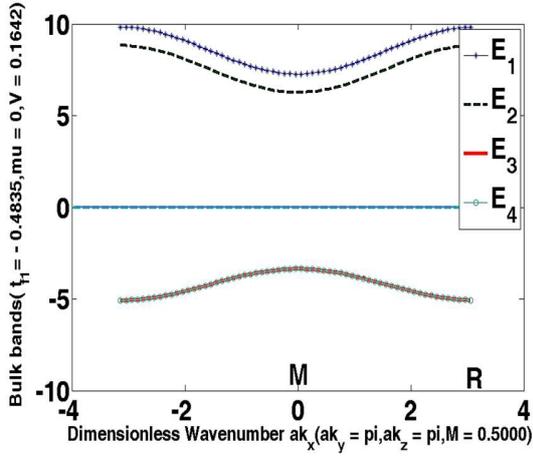
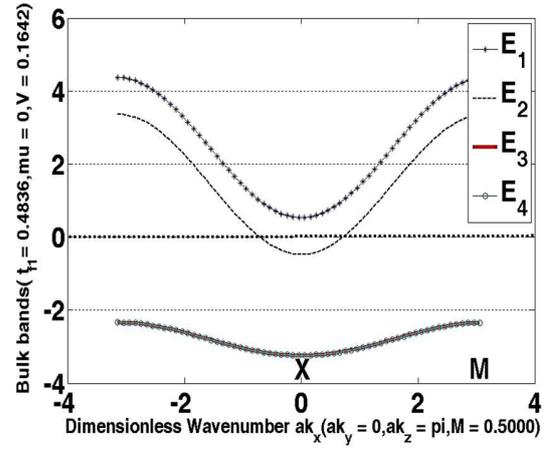

(c)

(d)t

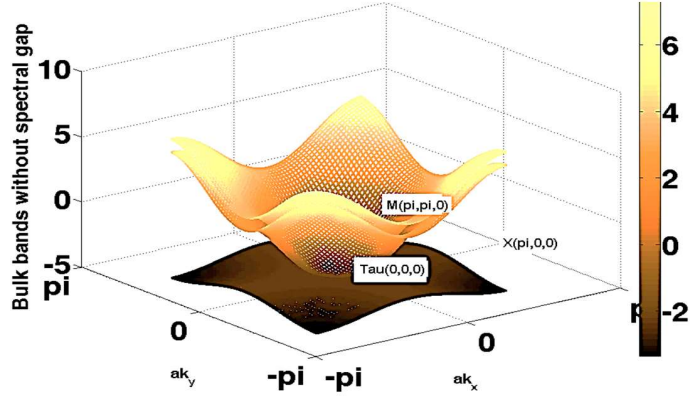

(e)

**Figure 2.** (a) A 3D diagrammatic representation of the four bands given by Eq.(4) with gap opening due to $t_{f_1} < 0$, and $M \neq 0$. **(b) and (c)** The representation of the same four bands along the paths $\bar{M}(-\pi, 0, \pi) - X(0,0,\pi)$-- $M(\pi, 0, \pi)$ and $\bar{R}(-\pi, \pi, \pi) - M(0, \pi, \pi)$-- $R(\pi, \pi, \pi)$, respectively, with $t_{f_1} < 0$, and $M \neq 0$. The band E₃ and E₄ are nearly degenerate. **(d)** The plot of the four bands along the path $\bar{M}(-\pi, 0, \pi) - X(0,0,\pi)$-- $M(\pi, 0, \pi)$ for metallic bulk case($t_{f_1} > 0$), and $M \neq 0$. Once again the band E₃ and E₄ are nearly degenerate. **(e)** A 3D representation of the four bands for $t_{f_1} > 0$. The numerical values of the parameters used in the plots are $t_{d_1} = 1.0000$, $|t_{f_1}| = 0.4835$, $t_{d_2} = 0.1$, $t_{f_2} = 0.1$, $t_{d_3} = 0.01$, $t_{f_3} = 0.01$, $\epsilon_f = -0.02$, V= 0.1642, μ = 0, $|b|^2 = 0.83$, and $M = 0.5000$. The Fermi energy is represented by the horizontal line.

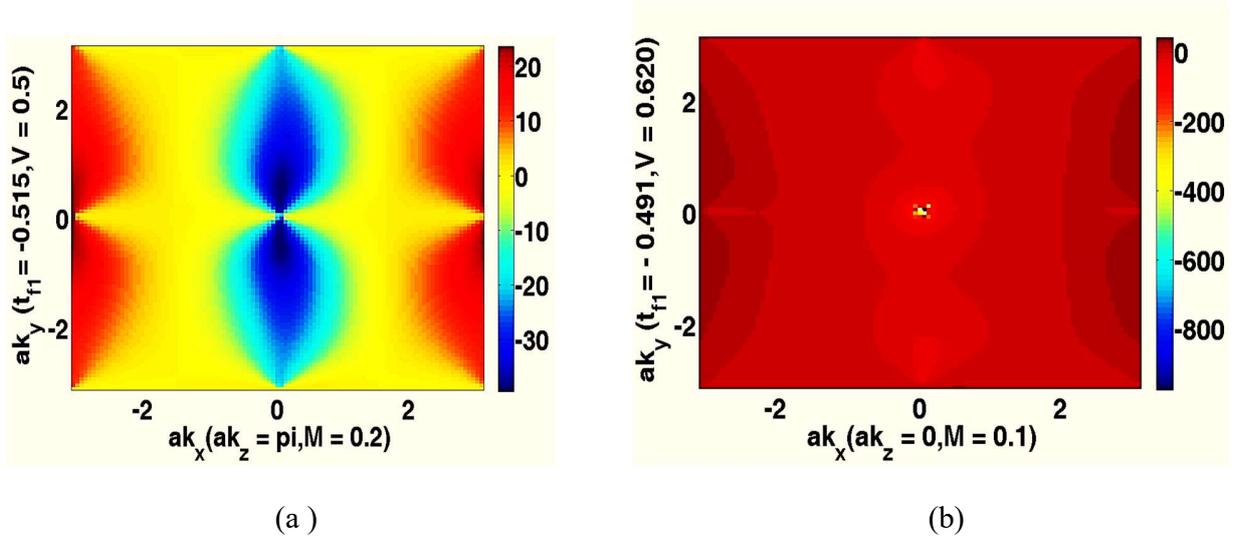

(a)  (b)

**Figure 3.** The contour plots of the Berry-curvature in the z-direction as a function of the dimension-less wave vector components $ak_x$ and $ak_y$. The numerical values of the parameters used in the plots in (a) and (b) are $t_{d_1} = 1$, $t_{f_1} = -0.5$, $t_{d_2} = 0.1$, $t_{f_2} = 0.1$, $t_{d_3} = 0.01$, $t_{f_3} = 0.01$, $\epsilon_f = -0.02$, $|b|^2 = 0.83$, and $\mu = 0$. The additional parameter values in **(a)** $ak_z = \pi$, $t_{f_1} = -0.515$, V = 0.50, and M = 0.20, while in **(b)** $ak_z = 0$, $t_{f_1} = -0.491$, V = 0.620, and M = 0.10. The chern number calculated in the two cases are **(a)** $\mathcal{C} = -2.0043 \sim -2$, and **(b)** $\mathcal{C} = 2.0073 \sim +2$.

For surface state (Hamiltonian) near the $\bar{X}$ point, one can use the same method but change the $z$ to the $x$ direction. The Pauli matrices $\boldsymbol{\sigma}$ and $\boldsymbol{\tau}$ are acting in the space of bands and fulfil the relation $\Theta\ \boldsymbol{\sigma}/\boldsymbol{\tau}\ \Theta^{-1} = -\ \boldsymbol{\sigma}/\boldsymbol{\tau}$, where $\Theta$ is the time reversal operator. Moreover, since $\Theta$ acts as complex conjugation only in the position basis and not on in any other basis, the Hamiltonian (8) is time reversal invariant.

The eigenvalues $E_n (n = 5,6,7,8)$ of $h_{\text{surface}}(k,\mu,|b|,q)$ in (8) is given by a quartic $E_n^4 + a\ E_n^3 + \mathfrak{b}\ E_n^2 + c\ E_n + d = 0$. The coefficients $(a, \mathfrak{b}, c, d)$ are given by $a = -2(\widetilde{E_k^d}(\mu,k) + \widetilde{E_k^f}(\mu,|b|,k))$, $\mathfrak{b} = \{(\widetilde{E_k^d}(\mu,k) + \widetilde{E_k^f}(\mu,|b|,k))^2 - M^2 + 2C^2\}, C^2 = (\widetilde{E_k^d}(\mu,k) \times \widetilde{E_k^f}(\mu,|b|,k)) - \vartheta_{z0}^2 + (\vartheta_x^2 + \vartheta_y^2), c = \left\{M^2\widetilde{E_k^f}(\mu,|b|,k) - C^2(\widetilde{E_k^d}(\mu,k) + \widetilde{E_k^f}(\mu,k,|b|))\right\}, \quad d = -M^2 \times \widetilde{E_k^f}^2(\mu,|b|,k) + C^4 - 4\ \vartheta_{z0}^2(\vartheta_x^2 + \vartheta_y^2)$. In view of the Ferrari's solution of a quartic equation, one obtains the roots as

$$E_n(s,\sigma,k,|b|) = \sigma\sqrt{\frac{\eta_0(k)}{2} - \frac{a}{4}} + s\left(b_0(k) - \left(\frac{\eta_0(k)}{2}\right) + \sigma\ c_0(k)\sqrt{\frac{2}{\eta_0(k)}}\right)^{\frac{1}{2}}, \quad (10)$$

where $\sigma = \pm 1$ is the spin index and $s = \pm 1$ is the band-index. The functions appearing in Eq. (10) are given by

$$\eta_0(k) = \frac{2b_0(k)}{3} + (\Delta(k) - \Delta_0(k))^{\frac{1}{3}} - (\Delta(k) + \Delta_0(k))^{\frac{1}{3}},\ \Delta_0(k) = \left(\frac{b_0^3(k)}{27} - \frac{b_0(k)d_0(k)}{3} - c_0^2(k)\right), \quad (11)$$

$$\Delta(k) = \left(\frac{2}{729}b_0^6 + \frac{4d_0^2 b_0^2}{27} + c_0^4 - \frac{d_0 b_0^4}{81} - \frac{2b_0^3}{27} + \frac{2c_0^2 b_0 d_0}{3} + \frac{d_0^3}{27}\right)^{1/2},\ b_0(k) = \left\{\frac{3a^2 - 8\mathfrak{b}}{16}\right\}, \quad (12)$$

$$c_0(k) = \left\{\frac{-a^3 + 4a\mathfrak{b} - 8c}{32}\right\},\ d_0(k) = \frac{-3a^4 + 256d - 64ac + 16\ ^2\mathfrak{b}}{256}. \quad (13)$$

The surface state energy spectra (SSES) $E_n(n = 5,6,7,8)$, given by Eq. (10), are plotted in Figure 4 as function of the dimensionless wave vector components. A 3D representation of the four bands, involving degeneracy and showing no spectral gap at $\bar{\Gamma}$ (0,0) appears in Figure 4(a). The curves in 4(b) correspond to the 2D representation of (a). The numerical values of the parameters used in all the plots are $t_{d_1} = 1.00$, $|t_{f_1}| = 0.50$, $t_{d_2} = 0.1$, $t_{f_2} = 0.01$, $t_{d_3} = 0.001$, $t_{f_3} = 0.001$, $\epsilon_f = -0.02$, V= 0.20, $|b| = 0.91$, and $M = 0$. In both the figures, there is band degeneracy as $M = 0$. The hopping parameter $t_{f_1}$ is negative and, therefore, the figures correspond to the insulating bulk. The presence of partially empty conduction band in Figure 4(b) clearly indicates the presence of conducting surface state. This is in conformity with the previous experimental observation concerning the de Haas-van Alphen (dHvA)effect in SmB$_6$ **[48,49]**. The Dirac point ( DP) – like feature of surface state excitation spectrum connecting the valence band with the conduction band has been reported in earlier **[50]**. It was shown by the authors that the surface states involve three Dirac cones. One of them is located at $\bar{\Gamma}$ (0,0)point and the other two are located at two $\bar{X}$ points. Here the band touching occurs at $\bar{\Gamma}$ (0,0) point without the explicit Dirac point feature. The reason for the absence is not far to seek. It is due to the non-linearization of the Hamiltonian matrix around $\bar{\Gamma}$ and $\bar{X}$ points. It must be mentioned in passing that non-linear structure in the figure 4(b) hints at the possibility of hosting massive Dirac fermions.

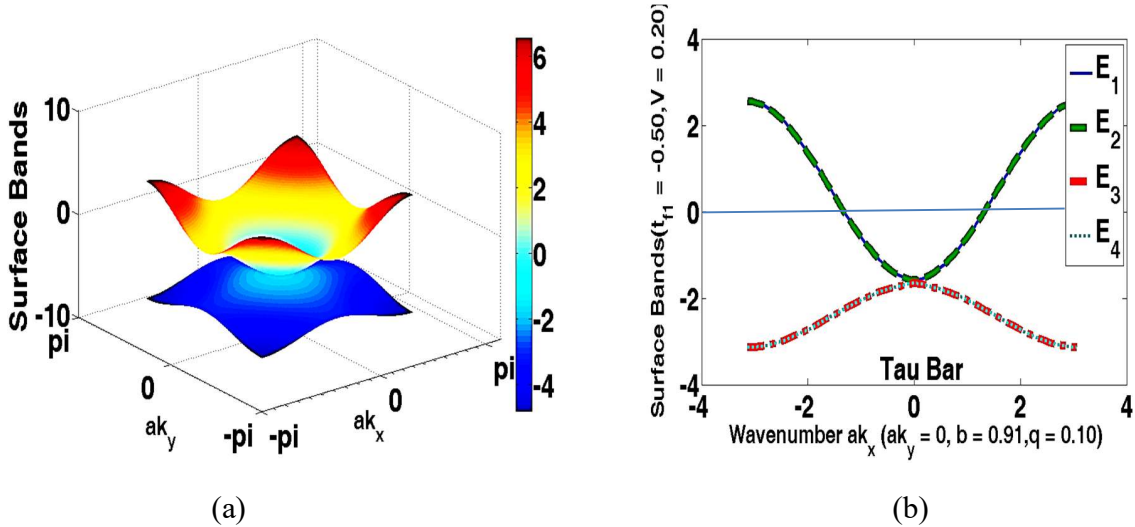

(a) (b)

**Figure 4.** The plots of the surface band energies (Eq. (10)) when the bulk is insulator. The Figure (a) corresponds to 3D plots of the four bands, whereas (b) to its 2D counterpart. The band touching occurs at $\bar{\Gamma}$ (0,0) point without the explicit Dirac point feature. The numerical values of the parameters used in both the plots are $t_{d_1} = 1.0000$, $|t_{f_1}| = 0.50$, $t_{d_2} = 0.01$, $t_{f_2} = 0.01$, $t_{d_3} = 0.001$, $t_{f_3} = 0.001$, $\epsilon_f = -0.02$, V= 0.20, $|b|^2 = 0.91$, and $M = 0$. The horizontal solid line represents the Chemical potential μ = 0.

The Dirac-point (DP) feature was observed in several experiments, viz. by scanning tunneling microscopy **[51,52]**, angle-resolved photoemission spectroscopy (ARPES) **[53,54]**, the circular dichroism ARPES **[55]**, and so on. In order to obtain DP here, the long- wavelength or low-energy limit of the Hamiltonian (8) needs to be considered. For this purpose, the Taylor series-based replacements

$$\sin(ak_j) \to ak_j + O(a^3 k_j^3), \cos(ak_j) \to (1 - (\tfrac{1}{2}a^2 k_j^2) + O(a^4 k_j^4)) \qquad (14)$$

are necessary ($j = (x, y)$) around $\bar{\Gamma}$ point. For a linearized surface state (Hamiltonian) near the $\bar{X}$ point, one can use the same method but change the $z$ to the $x$ direction. One gets access to almost DP like feature around $\bar{\Gamma}$ and $\bar{X}$ points as shown in Figure 5(a). The figure displays the band inversion as well. In Figure 5(b), 3D representation of the two surface bands close to the Fermi energy have been shown. The parameter values used are $t_{d_1} = 1$, $t_{f_1} = -0.80$, $t_{d_2} = 0.01$, $t_{f_2} = 0.01$, $t_{d_3} = 0.001$, $t_{f_3} = 0.001$, $\epsilon_f = -0.02$, V = 0.10, $q = 0.10$, $\mu = 0$, $|b| = 0.90$, and $M = 0$. The representations are in agreement with the Dirac cone topological surface states calculated by DFT [50]. A plot of $|b|$ as a function of $R$ (see Eq.(2)) in the Figure 5(c) is shown below. The reason is to justify the value $|b| = 0.90$ used in Figures 5(a) and 5(b) to obtain the satisfactory graphical representation. This value of $|b|$ corresponds to $R \sim 250$. Thus, the conclusion from the graphical representations is that the strong *f*-electron correlation seems to lead to large effective mass of the carriers [40]. The DP-like feature is in favor of the fact that the surface state is topologically non-trivial. Further evidence of the non-triviality will be sought by the analysis given below with a part of the surface BZ around $\bar{\Gamma}$ point. The similar discussion could be made for the case around $\bar{X}$ point.

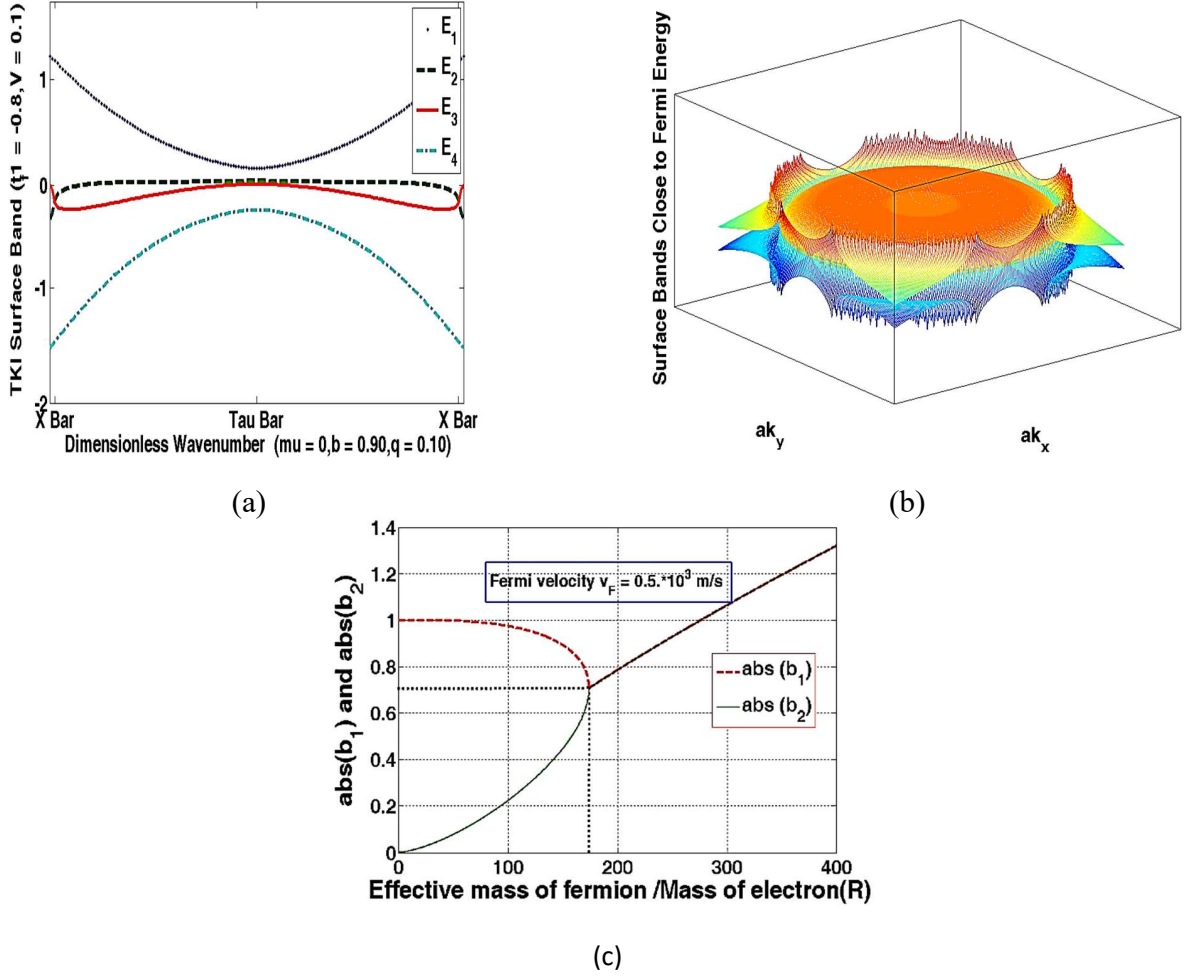

**Figure 5.(a)** The 2D representation of the four-band spectrum with the DP feature at $\bar{\Gamma}$ and $\bar{X}$ points. **(b)** The (3D) plot of the two surface bands close to the Fermi energy. The numerical values of the parameters used in the plots are $t_{d_1} = 1$, $t_{f_1} = -0.80$, $t_{d_2} = 0.01$, $t_{f_2} = 0.01$, $t_{d_3} = 0.001$, $t_{f_3} = 0.001$, $\epsilon_f = -0.02$, V = 0.10, $q = 0.10$, $\mu = 0$, $|b| = 0.90$, and $M = 0$. **(c)** A plot of absolute values of $b(|b_1|$ and $|b_2|$) as function of R for $v_F = 0.5 \times 10^3 \frac{m}{s}$.

The appearance of topologically-protected surface states is the physical consequence of the nontriviality. The time reversal (TR) operator for a spin 1/2 particle is $\Theta = I^{2\times 2} \otimes \tau^y K$. The operator $K$ stands for the complex conjugation. The inversion symmetry (IS) operator, on the other hand, is constructed as $\Pi = I^{2\times 2} \otimes \tau^z$. The $\tau^j$ are Pauli matrices on two-dimensional spin space. The Hamiltonian under consideration, for $M = 0$, preserves the time reversal (TRS) and inversion symmetries (IS). It can be easily shown that $\langle \Theta\psi|\Theta\varphi\rangle = \langle\varphi|\psi\rangle$ taking eigenstate of the z-component of the spin operator $I^{2\times 2}\otimes\tau^z$ as the basis. Also, $\Theta\gamma^0\Theta^{-1} = \gamma^0$, $\Theta\gamma^1\Theta^{-1} = -\gamma^1$, and $\Theta\gamma^j\Theta^{-1} = \gamma^j$, where $j = 2,3$. Similarly, $\Pi\gamma^0\Pi^{-1} = \gamma^0$, $\Pi\gamma^j\Pi^{-1} = -\gamma^j$ ($j = 1,2$), and $\Pi\gamma^k\Pi^{-1} = \gamma^k$ ($k = 3,5$). Since only $\gamma^0$, and $\gamma^3$ are even under time reversal and inversion, at a time reversal invariant momentum(TRIM) $K_i$ where the system preserves both TR and IS, the surface Hamiltonian (8) will have the following form:

$$h_{\text{surface}}(k = K_i, M = 0) = [\frac{\widetilde{E_k^d}(\mu,k)+\widetilde{E_k^f}(\mu,|b|,k)}{2}\mathbb{I} + \frac{\widetilde{E_k^d}(\mu,k)-\widetilde{E_k^f}(\mu,|b|,k)}{2}\gamma^0 + \vartheta_{z0}\gamma^0\gamma^3]. \quad (15)$$

The eigenvalues of $\Pi$, $\gamma^0$ and $\gamma^0\gamma^3$ are $\pm 1$ (multiplicity 2). The eigenvectors corresponding to the eigenvalues $+1$ and $-1$ of the parity operator, respectively, are $|+\rangle = (1/\sqrt{2})(1\ 0\ 1\ 0)^T$ and $|-\rangle = (1/\sqrt{2})(0\ -1\ 0\ -1)^T$. These are the relevant eigenstates. It is easy to see that

$$\langle+|h_{\text{surface}}(k = K_i, M = 0)|+\rangle = \widetilde{E_k^d}(\mu,k) + \vartheta_{z0} = E_+, \quad (16a)$$

$$\langle-|h_{\text{surface}}(k = K_i, M = 0)|-\rangle = \widetilde{E_k^d}(\mu,k) - \vartheta_{z0} = E_-. \quad (16b)$$

Obviously enough, $E_- < E_+$. Thus, the occupied state corresponds to the parity eigenvalue $-1$. Since the $Z_2$ invariant is determined by the parity eigenvalue of the occupied state, the outcome that emerges from the investigation around $\bar{\Gamma}$ point is that one has the surface state non-triviality (when $M = 0$) on hand. The finding does agree with the theoretical and experimental observations [2,50-56] reported earlier. The work in ref.[57], in particular, reports the first direct observation of spin-textured non-trivial surface states for the compound $SmB_6$ at the (111) surface different from the natural (100) cleavage plane.

## 4. Results and Discussion

As already stated, the $SmB_6$ TKI system is quite mysterious [4-16]. Despite this, the analysis presented could indicate that the model is likely to correspond to a non-trivial topological insulator in the absence of FM. The magnetic impurities introduced open a large exchange gap at the Dirac point (cf. Figures 2(b)); the chemical potential lies within the gap. This is an important requirement for the chern number to have integer values apart from the broken TRS. The chern number plays the role of the topological invariant of the quantum Hall system. One may add that this does not give us carte blanche to declare that QAH effect is definitely observable in the system envisaged as the effect has only been found in the sub- kelvin range **[58-60]** hitherto. Also, it needs to be mentioned that deciphering the mystery of the triviality/ non-triviality does not get over by decoding only a special case. The conclusive evidence will emerge only when the entire BZ is examined. Furthermore, in order to obtain satisfactory bulk/surface band spectrum, it has been found that $|b|^2 \approx 0.90$. The corresponding value of the effective mass is much greater than $m_e$ where $m_e$ is the bare mass of an electron. This is in agreement with the theoretical and experimental finding **[2,40]**. The indication is that the signature of the strong correlation is the large effective mass of the spin-momentum locked carriers of the system **[40]**. These points constitute the significance of the present study.

There is, however, an issue which concerns the fact that only the lowest-order cubic harmonics **[56]** in HT in (1) has been taken into account. In fact, it is desirable to introduce better odd-parity expressions for this form factor and re-investigate the present problem from a better perspective bringing about some improvement in the surface Hamiltonian. This is expected to facilitate more refined analysis of surface state excitation spectrum including DC feature. Another issue is the heuristic evanescent wave approach made to obtain the surface state Hamiltonian following ref.**[44]**. Instead of this, one may consider a slab geometry for this purpose with the thickness along the z direction. The thickness may be limited in z ∈ [−d/2, d/2], where d is measure of the penetration depth of surface states. One may further assume the open/ non-open boundary conditions and investigate the existence / non-existence of surface states. These are some of the essential points which need to be looked into for further refinement.

As regards the scope of the future work related to this paper, it is important to note that the Kondo problem **[61]** provides a paradigm for a variety of physical effects **[61-64]** involving strong electronic correlations. Local moment formation and Kondo screening are also a crucial ingredient of the Kondo physics. The scaling universality in the temperature dependence of physical quantities and their response to external fields at lower energies than a scaling energy (Kondo temperature $T_k$) is the central feature of the Kondo effect. The Kondo temperature is usually determined by the spin susceptibility $\chi_s = \mu_B \frac{\partial \langle m_d \rangle}{\partial B}|_{B=0, T=0}$ or the *T*-linear specific heat coefficient $\gamma_d = \text{limit}_{T \to 0}(\frac{S(T)}{T})$. Here, $m_d$ is the magnetization, and $S(T)$ is the specific heat. Upon treating the compound $SmB_6$ as the platform, the scaling universality may be explored. The temperature–bias–driven spin thermo-current and the spin susceptibility of the compound presently using the methodology presented in the paper is the ongoing investigation. These response functions are expected to show a universal Kondo scaling as a function of $\frac{T}{T_k}$.

The spin-polarized ARPES measurements **[55]** confirm the surface helical spin texture. Be that as it may, the Rashba spin-orbit coupling (RSOC) can arise in a system due to the proximity of material lacking in the structural inversion symmetry (SIS). It may also arise due to the impurity-induced structural distortion **[65]**. It would be, therefore, interesting to see how does surface state react to Rashba splitting **[24]**. This problem needs an extensive investigation perhaps introducing a term representing RSOC between the *d*-electrons, viz.

$$h_z = \left[\left(\frac{1}{2}\right)\left(-\alpha_0 \sin(k_y a)\right) \sigma_x \otimes (\tau_z + \tau_0) + \left(\frac{1}{2}\right) \alpha_0 \sin(k_x a) \sigma_y \otimes (\tau_z + \tau_0)\right], \quad (17)$$

in Eq.(7). Here $\alpha_0$ stands for the strength of RC. There are many other complications **[5,6,8, 66-70],** such as the unusual dichotomy between de Haas-van Alphen( dHvA) **[48,49]** and Shubnikov-de Haas (SdH) quantum oscillations **[68],** to bring home the point that the system needs much deeper and concerted scrutiny.

It is reported here that the realization QAH insulators with Chern number C = ±2 is possible under zero magnetic field. It will be shown in a separate communication that the Chern number in the same sample configuration can be tuned up to C = 4 by varying the magnetic doping concentration. The realization of such tunable chern number insulators could be a platform for low-power-consumption electronics **[71]**. The material is also useful for spintronics and quantum computation applications **[72]**.

Moreover, the effect of rise in temperature on the stability of QAH phase discussed here is a critical issue. The reason being only at sufficiently low temperature $(T)$, $k_B T$ becomes smaller than the energy difference between discrete quantum states leading to unfolding of quantum effects. In a bid to obviate this concern, a low-energy time-periodic, surface Hamiltonian, which is a variant of the present extended PAM obtained using the Floquet theory **[73-79],** is being theoretically investigated upon. In the high-temperature correlated metallic phase, the system surface is irradiated by the circularly polarized electro-magne tic radiation in the Floquet-Magnus limit **[75]**. Interestingly, the radiation field leads to the possibility of the emergence of the quantum anomalous Hall (QAH)state with the integer values of the chern number. As regards the temperature dependence of the anomalous Hall effect, recently this has been observed in $HgCr_2Se_4$ **[80].** A detailed investigation is needed to show a similar effect in the case of $SmB_6$ with FM impurities.

There is a proposal **[69]** for the possibility of nodal semi-metallic behavior in $SmB_6$ due to the electron-impurity scattering. In fact, it was shown that when inverse lifetime of QPs become comparable to the hybridization parameter, the system undergoes a topological crossover to an asymmetric nodal semi-metal phase. To treat this aspect on the line of the present framework, the energy bands in Eq.(4) may be replaced by

$$E_n(s,\tau,k,|b|,M,V) = \frac{(\widetilde{E_k^d}(\mu) + \widetilde{E_k^f}(\mu,|b|) + \tau M - i\Gamma_n(k))}{2}$$

$$+ s \left[ \frac{(\widetilde{E_k^d}(\mu) - \widetilde{E_k^f}(\mu,|b|) + \tau M - i\Gamma_n(k))^2}{4} + \vartheta_x^2 + \vartheta_y^2 + \vartheta_z^2 \right]^{\frac{1}{2}} \quad (18)$$

where $\Gamma_n(k)$ represents the energy level broadening given by ($\hbar v_0 / \beta$), $\beta$ is a semiclassical mean free path, $v_0 = \sqrt{\sum_j v_j^2}$, $v_j = \hbar^{-1}(\frac{\partial E_n(s,\tau,k,|b|,M,V=0)}{\partial k_j})$, and $j = (x,y,z)$.

## 5. Conclusions

The Hermicity of an open system is universally lost as the system always involves certain degrees of gain and loss. This is a very contentious issue. The ubiquitous electron-electron, electron-impurity, and electron-phonon scatterings in an electronic system are responsible for this loss/gain and give rise to quasiparticles (QPs)with finite lifetime. Effort will be made in future to look into this aspect broadly along the line of the idea represented in Eq.(18).

In conclusion, it is perhaps difficult to achieve enhancement in the current understanding of strongly correlated topological insulators unless other TKI candidates are discovered and thoroughly studied. Looking at the controversies and the possibilities **[5,6,8, 66-70]**, it is anybody's guess that there are many unsettled issues. It is expected that such issues will motivate the condensed matter physics aficionado to delve deeper into the problems of this mysterious compound.

## Appendix A

The bulk eigenstates corresponding to the eigenvalues in (4) are given by

$$|u_n(k_x, k_y, k_z)\rangle = \frac{1}{N_n^{1/2}} \begin{pmatrix} 1 \\ -\frac{\vartheta_z}{A} \\ \frac{2M\,\vartheta_z}{B_n(k)} \\ \frac{-C_n(k)}{AB_n(k)} \end{pmatrix}, \quad n=1,2,3,4, \; A=(\vartheta_x - i\vartheta_y) \qquad (A.1)$$

$$A = \vartheta_x - i\vartheta_y, \; B_n(k) = (\vartheta_x^2 + \vartheta_y^2 + \vartheta_z^2) - (E_n - \widetilde{E_k^d}(\mu,k) + M)(E_n - \widetilde{E_k^f}(\mu,k)), \quad (A.2)$$

$$C_n(k) = \left(E_n - \widetilde{E_k^d}(\mu,k) + M\right)\vartheta_z^2 + \left(E_n - \widetilde{E_k^d}(\mu,k) - M\right)(\vartheta_x^2 + \vartheta_y^2) + ((E_n - \widetilde{E_k^d}(\mu,k))^2 - M^2)(E_n - \widetilde{E_k^f}(\mu,k)), \qquad (A.3)$$

$$N_n = \left(1 + \frac{\vartheta_z^2}{A^*A} + \frac{4M^2\vartheta_z^2}{B_n^2(k)} + \frac{C_n^2(k)}{A^*AB_n^2(k)}\right). \qquad (A.4)$$

In view of (8), in order to obtain the surface state $|u_n(k_x, k_y)\rangle$, one needs to replace $\vartheta_z$ by $\vartheta_{z0}$ in Eq.(A.1).